\documentclass[a4paper,11pt]{article}
\usepackage{amsfonts,amssymb}

\usepackage{graphicx}
\usepackage{amsmath}

\newtheorem{theorem}{Theorem}

\newtheorem{corollary}[theorem]{Corollary}

\newtheorem{definition}[theorem]{Definition}
\newtheorem{example}[theorem]{Example}

\newtheorem{lemma}[theorem]{Lemma}

\newtheorem{remark}[theorem]{Remark}

\DeclareMathOperator{\expect}{{\mathbb E}}
\DeclareMathOperator{\prob}{{\mathbb P}}
\DeclareMathOperator{\real }{Re}
\DeclareMathOperator{\imag}{Im}
\DeclareMathOperator{\rot}{rot}
\DeclareMathOperator{\tr}{Tr}

\newcommand{\half}{\frac{1}{2}}
\newcommand{\qed}{$\square$}

\title{The Gibbs ensemble of a vortex filament
}
\author{
  \\
  { Franco Flandoli }              \\
  {\small\it Dipartimento di Matematica Applicata} \\ 
  {\small\it Universit\`{a} di Pisa, via Bonanno 25 bis} \\
  {\small\it 56126 Pisa, Italy}          \\
  {\small\tt flandoli@dma.unipi.it} \\
  \and \\
  { Massimiliano Gubinelli~\footnote{Present address: studio \#96, Scuola
      Normale Superiore, P.za dei Cavalieri, 7, 56126  Pisa, Italy} }              \\
  {\small\it Dipartimento di Fisica and INFN -- Sezione di Pisa}    \\
  {\small\it Universit\`a di Pisa, via Buonarroti 2, Ed. B}        \\
  {\small\it 56127 Pisa, Italy}          \\
  {\small\tt mgubi@cibs.sns.it} \\
}

\date{July 2000}

\begin{document}

\maketitle

\begin{abstract}
We introduce a statistical ensemble for a single vortex
filament of a three dimensional incompressible fluid. 
The core of the vortex is modelled by a quite generic
stochastic process. We prove the existence of the partition function for
both positive and a limited range of negative temperatures.   
\end{abstract}

\section{Introduction}
Certain investigations of turbulent 3-D fluids seem to indicate that the
vorticity field of the fluid is strongly concentrated along thin structures,
called vortex filaments. A. Chorin has developed a statistical-mechanics
theory of vortex filaments to describe the statistics of turbulent flows,
see \cite{Ch}. In his investigations vortex filaments are modeled by
trajectories of self-avoiding walks on a lattice. 
The kinematic energy $H$ of such
filaments is properly defined (due to the lattice cut-off) and Gibbs
measures of the form 
\begin{equation*}
d\mu _{\beta }=\frac{1}{Z_{\beta }}e^{-\beta H}dP_{SAW}
\end{equation*}
are carefully analyzed, where $P_{SAW}$ denotes the self avoiding walk
measure. Both positive and negative temperatures are considered, similarly to
Onsager 2-D theory of point vortices.

For many reasons it is interesting to attempt a similar description of
vortex filaments by means of continuous curves, like trajectories of
Brownian motion or other stochastic processes, not restricted to a lattice.
An informal proposal has been given by G. Gallavotti \cite{Ga}, section
II.11, and an extensive rigorous approach has been developed by P.L. Lions
and A. Majda \cite{PLL - M} under a specific idealization, called nearly
parallel vortices. The outstanding work \cite{PLL - M} yet contains the
basic restriction that the filaments cannot fold, while folding is a major
feature of general vortex filaments, necessary to prevent energy increase as
a consequence of vortex stretching (see \cite{Ch}, Ch. 5). In the
approximation of \cite{PLL - M} the definition of energy and Gibbs measures
is not a basic difficulty, while the aim is to reach a mean field result and
several effective characterizations of the mean field distribution. The
Wiener measure arises naturally in \cite{PLL - M} as \ a reference measure.
Only positive temperatures have been considered in this approach.

In \cite{Fl2} another model of vortex filaments based on 3-D Brownian paths has
been introduced. The attempt of \cite{Fl2} has been to keep into account the
full energy, without any cut-off or idealization, in particular allowing for
vortex folding. But the full energy of a single vortex curve is infinite (a
well known fact for smooth curves, true also for Brownian curves). Therefore
it is necessary to consider vortex structures with a certain cross section
instead of a single curve. A cross section exists in physical vortex
structures, and seems to be fractal. The objects introduced in \cite{Fl2}
are thus vortex structures with a Brownian core and a fractal cross section.
In \cite{Fl2} the energy of such filaments is rigorously defined, proved to
be finite with probability one, with finite moments of all orders, and a
relation with the intersection local time of Brownian motion is established.

Among the many questions left open by \cite{Fl2} there is the existence of
the Gibbs measures, i.e. the exponential integrability of the energy with
respect to the Wiener measure. For positive temperatures this property would
be a consequence of the positivity of the energy, but the property $H\geq0$
is not clear form the approach of \cite{Fl2}.

The aim of this paper is to give a new definition of the energy $H$ for the
Brownian filaments of \cite{Fl2} (and for more general core processes),
which in particular shows that 
\begin{equation*}
H\geq 0.
\end{equation*}
This implies that the Gibbs measure 
\begin{equation*}
d\mu _{\beta }=\frac{1}{Z_{\beta }}e^{-\beta H}d\prob
\end{equation*}
is well defined for all $\beta >0$, where $\prob$ is the Wiener measure (with
expectation $\expect$). In
addition, we prove that 
\begin{equation*}
\expect e^{-\beta H}  <\infty \qquad \text{\ for sufficiently small 
\textit{negative} }\beta ,
\end{equation*}
i.e. the Gibbs measures are well defined also for high negative temperatures
(positive and negative temperatures meet at $|\beta^{-1}|=\infty $, see \cite{Ch}).
Finally, we show that 
\begin{equation*}
\expect  e^{-\beta H}  =\infty \qquad \text{\ for sufficiently large
negative }\beta ,
\end{equation*}
so the range of negative temperatures that can be considered is restricted.
This phenomenon is similar to 2-D point vortices theory (see \cite{Ch}, \cite
{Mar-Pulv}), and to the case of the renormalized polymer measure, see~\cite{LeG3}.

The main difference between~\cite{Fl2} and the
present approach is that the energy in~\cite{Fl2} is expressed as a double
stochastic integral, while here it is the square of a single stochastic
integral (further integrated in some parameters). Finally, in~\cite{Fl2} the
filament core is Brownian, while here we consider more general processes
(the Brownian semimartingale), which include for instance also the Brownian
Bridge, a remarkable example since for closed paths the fields are
divergence free.

The energy $H$ is defined here by means of spectral analysis. This
representation leads to a formula for the energy spectrum in terms of
expectation of stochastic integrals. We also show how to derive the
representation of $H$ given in  \cite{Fl2} from the present one. The
intersection local time arises when $H$ is suitably decomposed as we
show in section~\ref{sec:otherexp}.
An aside comment: in definition~\ref{def_energy} below, we introduce a concept of
$\rho-$inertial energy of a stochastic process that can be applied to
a large variety of processes and perhaps it may be useful to
characterize certain fractal or regularity features.

The final section collects some remarks on the case of multiple vortex
filaments and on further developments of the theory. 
 
\section{The vortex filaments}

\subsection{Brownian semimartingales}

The \textit{core} of the vortex filaments introduced below is based on the
following class of processes. We call $3$-D Brownian semimartingale on a
filtered probability space $( \Omega ,\mathcal{A},( \mathcal{F}%
_{t}) _{t\in [ 0,T] },\prob) $ 
(with expectation $\expect$) any
stochastic process $\left( X_{t}\right) _{t\in \left[ 0,T\right] }$ on $%
\left( \Omega ,\mathcal{A},\prob\right) $, of the form 
\begin{equation}
X_{t}=W_{t}+\int_{0}^{t}b_{s}ds, \qquad t\in \left[ 0,T\right] ,
\label{Browsemim}
\end{equation}
where $( W_{t} ) $ is a $3$-D Brownian motion with respect to the
filtration $( \mathcal{F}_{t} ) _{t\in \left[ 0,T\right] }$ and $%
( b_{t} ) _{t\in \left[ 0,T\right] }$ is an $( \mathcal{F}%
_{t})$-progressively measurable process taking values in
$\mathbb{R}^{3}$, 
with at least the integrability property 
$$
\prob ( \int_{0}^{T}\|
b_{s}\| ds<\infty ) =1
$$
(we need stronger integrability properties below depending on the result we want to prove for
the energy).

Relevant examples of Brownian semimartingales are the Brownian motion itself
(considered in \cite{Fl2}), the solutions of stochastic differential
equations of the form 
\begin{equation*}
dX_{t}=f(t,X_{t})\, dt+dW_{t}
\end{equation*}
for suitable drift vector fields $f$, and the Brownian Bridge, recalled
below.

The relevant property of Brownian semimartingales used in the sequel is that
the quadratic variation of each component is $t$ and the mutual variations
between different components is zero, i.e. 
\begin{equation*}
\left[ X^{i},X^{j}\right] _{t}=\delta_{ij}t.
\end{equation*}
For the definition of the mutual variation (or bracket) $\left[ Y,Z\right]
_{t}$ between two real semimartingales, see for instance \cite{Ku} and \cite
{R-Y}.

\begin{example}
(Brownian Bridge) We follow \cite{R-Y}, in particular Exercise (3.18) of Ch.
IV. We take the time interval $\left[ 0,1\right] $, but it is a simple
exercise to rewrite the formulae in the general case. Given a $3$-D Brownian
motion $\left( B_{t}\right) _{t\in \left[ 0,1\right] }$, the Brownian Bridge 
$\left( X_{t} \right) _{t\in \left[ 0,1\right] }$ 
is defined as 
\begin{equation*}
X_{t} = B_{t}-tB_{1}.
\end{equation*}
We could work out many steps of our approach using this explicit expression,
but to unify the analysis we recall that $\left( X_{t} \right) $ is a
Brownian semimartingale with respect to a suitable filtration and Brownian
motion $\left( W_{t}\right) $. Precisely, on the probability space
$(\Omega, \mathcal{A}, \prob)$ where $(B_t)$ is defined, there exists a
filtration $(\mathcal{F}_t)_{t\in [0,1]}$ and a $\mathcal{F}_t$-Brownian motion
$(W_t)$ in $\mathbb{R}^3$ such that $(X_t)$ is adapted to
$(\mathcal{F}_t)$ and 
\begin{equation*}
X_{t} =W_{t}-\int_{0}^{t}\frac{X_{s}}{1-s}ds, 
\qquad t\in %
\left[ 0,1 \right] .
\end{equation*}
Since $X_{s} = \left( B_{s}-B_{1}\right) +\left(
1-s\right) B_{1}$, we have
\begin{equation*}
\expect \left\| \frac{X_{s}}{1-s}\right\| 
\leq C\left( 1+\frac{1}{\sqrt{1-s}}\right) 
\end{equation*}
so the process 
\begin{equation*}
b_{s}:=\frac{X_{s}}{s-1}
\end{equation*}
satisfies the assumption
\begin{equation*}
\expect \int_{0}^{1}\left\| b_{s}\right\| ds <\infty
.
\end{equation*}
Since $(b_s)$ is Gaussian, this implies stronger integrability
properties, as described in Example~\ref{example_of_brownian_bridge}. 
In particular, the Brownian
bridge satisfies the assumptions of Theorems~\ref{th1} and~\ref{th_finite1} below.

\end{example}

\subsection{Vortex filaments based on Brownian semimartingales}

Let $\left( X_{t}\right) _{t\in\left[ 0,T\right] }$ be a Brownian
semimartingale on a filtered probability space $( \Omega,\mathcal{A}%
,( \mathcal{F}_{t}) _{t\in [ 0,T] },\prob) $. We have
in mind that $\left( X_{t}\right) $ is the \textit{core} of the filament we
want to introduce, while the full filament is a collection of translates of $%
\left( X_{t}\right) $ of the form $\left( x+X_{t}\right) $ with $x$ varying
on a fractal set. This fractal set is in a sense the \textit{cross section}
of the filament. In reality the cross section has a complicate dependence on
the position along the filament, while in the idealized model proposed here
it is the same everywhere. The only realistic feature we try to keep into
account is the fractality of the cross section with respect to artificial
smoothings like a tubular neighborhood of the core (e.g. a Brownian sausage).

Instead of specifying the fractal cross section at the set-theoretical
level, it is more convenient to describe it by means of a finite
measure $\rho$ (supported on the fractal set) on the Borel sets of $%
\mathbb{R}^{3}$. We thus consider a distributional vorticity field of the
form 
\begin{equation}
\xi( dx) = \int\rho\left( dy\right) \int_{0}^{T}\delta\left(
dx-\left( y+X_{t}\right) \right) \circ dX_{t}  \label{vort}
\end{equation}
where $\circ dX_{t}$
denote, as usual, the Stratonovich integration with respect to the semimartingale $\left(
X_{t}\right) $. To have an intuitive understanding of this definition it is
helpful to consider first the ideal case of a vorticity field concentrated
over a smooth curve $\left( \gamma\left( t\right) \right) _{t\in\left[ 0,T%
\right] }$ in $\mathbb{R}^{3}$ of unit intensity. 
Its formal definition would be 
\begin{equation*}
\xi\left( dx\right) = \int_{0}^{T}\delta\left( dx-\gamma\left( t\right)
\right) \,\dot{\gamma}\left( t\right) \,dt.
\end{equation*}
If we consider a collection of translates of $\gamma\left( t\right) $
weighted by a finite measure $\rho$ we find an expression like (\ref
{vort}) with $\gamma\left( t\right) $ in place of $X_{t}$. Passing from
smooth curves to the paths of a semimartingale, the natural operation is to
consider the Stratonovich integral. Another important motivation for the use
of Stratonovich integrals is the condition $\nabla \cdot \xi=0$, 
as explained in 
\cite{Fl2} (this condition is strictly satisfied only in the case of closed
or infinite filaments, and requires a Stratonovich integral in (\ref{vort})).

A cross section exists in reality, but to reduce the degrees of freedom of
the model it would be better in principle to use single curves. However,
both in the case of a smooth curve and a Brownian motion $X=W$, if we simply
take $\rho=\delta$ in (\ref{vort}) (i.e. we do not integrate in $\rho$), we
obtain later on an infinite energy, see \cite{Fl2}. This is the reason to
include in the model a cross section. The fractality of the cross section
(in contrast to easier tubular neighborhoods) is motivated by numerical
results (see for instance \cite{Bell-M}, \cite{Ch}). If we replace the
measure-theoretic description of the cross section
and use sets, we could say that we consider vortex structures of the form 
\begin{equation*}
\left\{ x+X_{t}\,;\,x\in\mathcal{A},\,t\in\left[ 0,T\right] \right\}
\end{equation*}
where $\mathcal{A}\subset\mathbb{R}^{3}$ is a compact set. This is the 
\textit{thickened} vortex structure.

With a minor effort in the subsequent analysis, 
we could take a random measure $\rho $
independent of $\left( X_{t}\right) $, with suitable integrability
properties. This generality is of interest to go in the direction of more
realistic models and in particular to have fields close to homogeneous and
isotropic. It is also important if one want to consider collections of
filaments. At the theoretical level it does not introduce any real novelty.

A rigorous definition of the random distribution (\ref{vort}) is not needed
below. Instead, still arguing at a formal level, let us introduce the
kinetic energy associated to $\xi $ (see \cite{Ch}, \cite{Fl2}): 
\begin{equation}
H=\frac{1}{8\pi }\int_{\mathbb{R}^{3}} \int_{\mathbb{R}^{3}} 
\left( \int_{0}^{T}\int_{0}^{T}\frac{1}{%
\left\| \left( x+X_{t}\right) -\left( y+X_{s}\right)
\right\| }\circ dX_{s}\circ dX_{t}\right) \rho \left( dx\right)
\rho \left( dy\right) .  \label{H}
\end{equation}
This is the natural expression from the general formula (meaningful for
fields with a certain regularity) 
\begin{equation}
H=\frac{1}{8\pi }\int_{\mathbb{R}^{3}}\int_{\mathbb{R}^{3}}\frac{\xi \left(
x\right) \cdot \xi \left( x^{\prime }\right) }{\left\| x-x^{\prime }\right\| }%
\,dxdx^{\prime }.
\label{eq:smoothenergy}
\end{equation}

A priori it is difficult to give a meaning to (\ref{H}), because of the
anticipating stochastic integrations and the singularities coming from the
denominator. A rigorous meaning has been given in \cite{Fl2} when $\left(
X_{t}\right) $ is a Brownian motion, and $\rho $ satisfies the condition 
\begin{equation}
\frac{1}{4\pi} \int_{\mathbb{R}^{3}} \int_{\mathbb{R}^{3}}
 \frac{1}{\left\| x-x^{\prime }\right\|} \rho \left(
dx\right) \rho \left( dx^{\prime }\right) <\infty .  \label{rho}
\end{equation}

The approach of~\cite{Fl2} is to rewrite~(\ref{H}) as a double It\^{o}
integral plus correction. In this work we present a completely different
approach in which, \emph{under the same main assumption}~(\ref{rho}), 
the energy is naturally written as a positive definite quadratic from. 
The easiest way to do this is to rewrite eq.~(\ref{H}) in spectral variables.
This is done in the next section where we present informal arguments to
bridge the gap between rigorous definition and the physical meaning
of the formulae.

\section{Spectral analysis}  
In this section we will perform formal calculation to justify the
subsequent rigorous results from the physical point of view. 
We start from the spectral decomposition of the random vorticity
distribution~(\ref{vort}) and obtain an expression for the energy
which can be shown to be well defined.

We use $| \cdot | $  to denote the absolute value of a complex number
and $\| \cdot \|$ for the Euclidean norm of vectors in $\mathbb{R}^3$ or
$\mathbb{C}^3 \sim \mathbb{R}^6$.

The Fourier transform of $\xi(dx)$ is
$$
\hat{\xi}(k) = \int_{\mathbb{R}^3} e^{-i k\cdot x} \xi(dx) =
\hat{\rho}(k) \int_0^T e^{-ik\cdot X_t} \circ dX_t
$$
where
$$
\hat{\rho}(k) = \int e^{ik\cdot x} \rho(dx).
$$

The corresponding velocity field of the fluid $u(x)$ under
appropriate conditions on its decay at infinity is
$$
u(x) = \frac{1}{4 \pi} \rot \int_{\mathbb{R}^3} \frac{\xi(dy)}{|x-y|}
$$
and has Fourier transform
$$
\hat{u}(k) = \int_{\mathbb{R}^3} e^{-i k\cdot x} u(x) dx = -
\frac{\hat{\rho}(k)}{\|k\|^2} \int_0^T e^{-ik\cdot X_t} \circ ik \wedge dX_t.
$$ 

The inertial energy of the fluid is  
\begin{equation}
\begin{split}
H & = \half \int_{\mathbb{R}^3} \|u(x)\|^2 dx = \frac{1}{2} \int_{\mathbb{R}^3} 
\|\hat{u}(k)\|^2 \frac{dk}{(2\pi)^3} 
\\ & =   \half \int_{\mathbb{R}^3} \frac{dk}{(2\pi)^3} \frac{|\hat{\rho}(k)|^2}{\|k\|^2} 
\left\| \frac{ik}{\|k\|} \wedge \int_0^T e^{ik\cdot X_t} \circ dX_t
\right\|^2
\\ & =   \half \int_{\mathbb{R}^3} \frac{dk}{(2\pi)^3} \frac{|\hat{\rho}(k)|^2}{\|k\|^2} 
\left\| \int_0^T e^{ik\cdot X_t} \circ p_k  dX_t
\right\|^2
\end{split}
\label{eq:inertial_energy}
\end{equation}
where $a \wedge b$ denotes the vector product in $\mathbb{R}^3$ and 
$$
p_k v \equiv v - \frac{k \otimes k}{\|k\|^2} v, \qquad v,k \in \mathbb{R}^3
$$
is the projection in $\mathbb{R}^3$ in the plane orthogonal to $k$. 

In the sequel we shall often deal with the Stratonovich integral 
$$
Y_{k,T} \equiv \int_{0}^{T}e^{ik\cdot X_{t}}\circ dX_{t}.
$$ 
Notice that it takes values in $%
\mathbb{C}^{3}$. It can be rewritten as 
\begin{equation}
\int_{0}^{T}e^{ik\cdot X_{t}}\circ dX_{t}=\int_{0}^{T}e^{ik\cdot
X_{t}}dX_{t}+\frac{ik}{2}\int_{0}^{T}e^{ik\cdot X_{t}}dt.  \label{strat1}
\end{equation}
and, given that
$$
ik\cdot Y_{k,T} = \int_0^T  e^{ik\cdot X_t} \circ ik \cdot dX_t =
e^{i k\cdot X_T} - e^{i k\cdot X_0}, 
$$
the following decomposition in components transverse and parallel
to $k$  holds
\begin{equation}
\begin{split}
Y_{k,T} & = p_k Y_{k,T} + \frac{k\otimes k}{\left\| k\right\|^2} Y_{k,T}
\\ & 
=\int_{0}^{T}e^{ik\cdot
X_{t}}p_{k}dX_{t}-\frac{ik}{\left\| k\right\| _{\mathbb{R}^{3}}^{2}}\left(
e^{ik\cdot X_{T}}-e^{ik\cdot X_{0}}\right).   \label{strat2}
\end{split}
\end{equation}

We observe that the relation
$$
p_k Y_{k,T} -  \int_{0}^{T}e^{ik\cdot
X_{t}}p_{k}dX_{t} = \frac{1}{2} [e^{ik\cdot X}, p_k X]_t = 0
$$
implies that it does not matter whether the
transverse part of~(\ref{strat2}) is defined as an It\^o or
Stratonovich integral. 

The definition of the energy for a vortex filament is, to some extent, not
unique. For example, taking the equation~(\ref{H}) and formally
computing its spectral representation we would obtain 
\begin{equation}
\tilde{H} \equiv \half \int_{\mathbb{R}^3} \frac{dk}{(2\pi)^3} \frac{|\hat{\rho}(k)|^2}{\|k\|^2} 
\left\| \int_0^T e^{i k \cdot X_t} \circ dX_t \right\|^2
\end{equation}
without projection of $Y_{k,T}$ onto the subspace transverse to $k$. 
The expression~(\ref{eq:smoothenergy}) is obtained under the assumption that the
vorticity field is divergenceless which means that the vortex filament
should be closed (i.e. $X_0 = X_T$) or should start and end at
infinity. Using~(\ref{strat2}) it is easy to see that that the
component of $Y_{k,T}$ parallel to $k$ is zero for closed paths so
that $\tilde{H} = H$. 
For finite and open paths we have instead
\begin{equation}
\tilde{H} - H = 2 \int_{\mathbb{R}^3}  \frac{dk}{(2\pi)^3} \frac{|\hat{\rho}(k)|^2}{\|k\|^4}
\sin^2\left(\frac{k \cdot (X_T-X_0)}{2}\right).
\label{eq:difference}
\end{equation}

\section{Definition of the energy}
The following definition of $\rho-$inertial energy of a stochastic
process $(X_t)$ may be of interest in itself, beside the aims of the
present paper. Notice that it is meaningful for a large class of
stochastic precesses including, besides the Brownian semimartingales
considered here, also every semimartingale and also many other classes
of processes. It is only needed that the martingale
$p_k Y_{k,T}$ be defined for every $k \in \mathbb{R}^3$. 
This is true for instance
for processes like those considered in~\cite{Bert},\cite{Zal}

\begin{definition}
\label{def_energy}
Given a probability measure $\rho $ on $( \mathbb{R}^{3},B( 
\mathbb{R}^{3})) $ and a semimartingale $( X_{t}) $,
we call $\rho $-\textit{inertial energy} of $( X_{t}) $ the
(possibly infinite) random variable 
\begin{equation}
H \equiv \int_{\mathbb{R}^3} d\nu(k) \left\|p_k Y_{k,T}\right\|^2
\label{defH}
\end{equation}
where
$$
d\nu(k) \equiv \half \frac{\left| \hat{\rho}\left( k\right) \right| ^{2}}
{\left\| k\right\|^{2}} \frac{dk}{(2\pi)^3}.
$$
Moreover it will be convenient to consider also the modified energy
\begin{equation}
\tilde{H} \equiv \int_{\mathbb{R}^3} d\nu(k) \left\| Y_{k,T}\right\|^2.
\label{defHtilde}
\end{equation}
\end{definition}
We have that $0 \le H \le \tilde{H} \le \infty$ and we
not exclude a priori that the $\rho $-inertial energy is infinite. For a
class of Brownian semimartingales we show now that both are finite with
probability one.

\begin{theorem}
\label{th1}
Let $( X_{t}) $ be a Brownian semimartingale of the form (\ref
{Browsemim}), with drift satisfying 
\begin{equation}
C_{b}\equiv \expect \left( \int_{0}^{T}\left\| b_{t}\right\| dt\right)^{2}<\infty .
\label{eq:bound_on_b}
\end{equation}
Let $\rho $ be a finite measure on $( \mathbb{R}^{3},B( 
\mathbb{R}^{3}) ) $ satisfying (\ref{rho}). 
Then the random variables $H$ and $\tilde{H}$ are finite with probability one,
\begin{equation}
\expect [H] \le \expect [\tilde{H}] <\infty .  \label{Hfinite},
\end{equation}
moreover, the difference
\begin{equation}
\tilde{H} - H  = 2 \int_{\mathbb{R}^3} \frac{dk}{(2\pi)^3} \frac{|\hat{\rho}(k)|^2}{\|k\|^4}
\sin^2\left(\frac{k \cdot (X_T-X_0)}{2}\right)
\label{eq:finitedifference}
\end{equation}
is finite a.s. and $\expect [\tilde{H}-H] < \infty$ for every $\rho$
of finite mass (the assumption~(\ref{rho}) is not required).
\end{theorem}

PROOF. $H$ and $\Tilde{H}$ are positive r.v., then  
if we prove 
$ \expect H < \infty$ and $\expect [H - \Tilde{H}] < \infty$ we have
$H \le \Tilde{H} <\infty$ with probability one. 

Consider eq.~(\ref{eq:difference}) and 
given an orthonormal basis in $\mathbb{R}^3$ with the
1-direction parallel to $X_T - X_0$ denote with $k_1$ the
projection of $k$ onto this direction, then  using  the
estimates $|\hat{\rho}(k)| \le |\hat{\rho}(0)|$, we have
\begin{equation}
\begin{split}
\tilde{H} - H & \le 2 \frac{|\hat{\rho}(0)|^2}{(2\pi)^3}
 \int_{\mathbb{R}^3} \frac{dk}{\|k\|^4}
\sin^2\left(\frac{k \cdot (X_T-X_0)}{2}\right)
\\ & = 2 \frac{|\hat{\rho}(0)|^2}{(2\pi)^3}
 \int_{\mathbb{R}^3}  \frac{  dk}{\|k\|^4} 
\sin^2\left( k_1 \frac{\|X_T-X_0\|}{2} \right) 
\\ & =  \frac{|\hat{\rho}(0)|^2}{(2\pi)^3} \|X_T-X_0\| 
 \int_{\mathbb{R}^3}  \frac{  dk}{\|k\|^4} 
\sin^2\left( k_1 \right)  \\ & \equiv D \|X_T-X_0\|
\end{split}
\label{eq:extraction_in_difference}
\end{equation}
where we rescaled out from the integral the factor $\|X_T - X_0\|/2$. 
Note that $D$ does not depend on the process $(X_t)$.
Since  $\expect \|X_T-X_0\| < \infty$ by
eq.~(\ref{Browsemim}) and  assumption~(\ref{eq:bound_on_b}), we have
that
$$
\expect [\tilde{H} - H] < \infty.
$$
  
The measure $d\nu$ is a finite measure, since its mass  
$$
\int_{\mathbb{R}^3} d\nu(k) = \frac{1}{8 \pi} \int \int \frac{1}{\left|
x-y\right|}\rho \left( dx\right) \rho \left( dy\right), 
$$
is finite by assumption. We have 
\begin{equation*}
\expect [H] =\int d\nu \left( k\right) \expect \left\| p_k Y_{k,T} \right\|^{2}
\end{equation*}
so the claim will be proved showing that the last expected value is
uniformly bounded in $k$: 
\begin{eqnarray*}
\expect \left\| p_k Y_{k,T}\right\|^{2}   
&\leq & 2 \expect \left\| \int_{0}^{T}e^{ik\cdot
    X_{t}} p_k dW_{t}\right\|^{2} 
+2\expect\left\| \int_{0}^{T}e^{ik\cdot
X_{t}} p_k b_{t}dt\right\|^{2}  \\
&\leq & 4T + 2\expect \left| \int_{0}^{T}\left\| e^{ik\cdot X_{t}} p_k
    b_{t}\right\|
 dt\right| ^{2}  \\
&\leq &4T+2\expect  \left| \int_{0}^{T}\left\|  b_{t}\right\| dt\right|
^{2} \\ &
= & 4T+2C_{b} < \infty.
\end{eqnarray*}
\qed

\subsection{The energy for a Brownian motion}
\label{sec:otherexp}
Our definition~(\ref{defHtilde}) for the modified energy is not substantially
different from the one given in~\cite{Fl2} in the case when $(X_t)$ is
a Brownian motion. This  explicitly shows the 
connection of the energy of the filament with the intersection local time of the core
process $(X_t)$. 

In the following we will need backward stochastic calculus (see~\cite{Ku} for
reference) with respect to
Brownian motion for which here we recall the basic setup. 
Let $(X_t)$ be a Brownian motion and introduce the $\sigma-$algebras
$$
\mathcal{F}^t_s = \sigma\left\{ W_u - W_v ; s \le v \le u \le t \right\}
$$
Given $t >0$, the family $(\mathcal{F}^t_s)_{s\in [0,t]}$ is a backward
filtration and the process $(\Tilde{X}^t_s)_{s\in [0,t]}$ defined as
$\Tilde{X}^t_s = X_t - X_s$ is a $\mathcal{F}^t_s-$adapted (backward)
Brownian motion. 

\begin{theorem}
\label{th:equivalence}
Let $(X_t)$ be a Brownian motion  then the
energy as defined in eq.~(\ref{defHtilde}) is equal to the energy defined
in~\cite{Fl2}, that is
\begin{equation}
\begin{split}
\tilde{H} & =  \int_0^T \!\int_0^T  G^\rho(X_t-X_s)\, \hat{d}X_s~dX_t + \\
& \qquad + \frac{1}{2} \int_0^T \left[G^\rho(X_T-X_t) + G^\rho(X_t-X_0)\right]\, dt \\
& \qquad + \frac{1}{4} \int_0^T\! \int_0^T (\rho * \rho)(X_t-X_s)\, dt~ds \\
& \qquad + \frac{1}{2} G^\rho(0) T \\
\end{split}
\label{equivalence}
\end{equation}
where 
$$
G^\rho(x) \equiv (\rho * G * \rho)(x) \equiv \int_{\mathbb{R}^3} \int_{\mathbb{R}^3} G(x-y-z)
\rho(dy) \rho(dz)
$$
and $G(x) \equiv - 1/4 \pi \|x\|$ and where the
measure $\rho * \rho$ is defined as
$$
(\rho * \rho)(f) \equiv \int_{\mathbb{R}^3} \int_{\mathbb{R}^3} f(x+y) \rho(dx) \rho(dy).
$$
\end{theorem}
\begin{remark}
The term 
$$
 \int_0^T\! \int_0^T (\rho * \rho)(X_t-X_s)\, dt~ds
$$
is the intersection local time (see e.g.~\cite{LeG1},\cite{Yo2})
$$
 \alpha(z, T) \equiv \int_0^T\! \int_0^T \delta(z+X_t-X_s)\, dt~ds
$$
integrated in $z$ with respect to $(\rho * \rho)$.
\end{remark}

PROOF. 
It\^o formula gives
\begin{equation*}
\tilde{W}_k  \equiv  \left\| Y_{k,T} \right\|^2 
 =  
2 \real \int_0^T \left( \int_0^t e^{-ik \cdot X_s} \circ dX_s \right) e^{ik \cdot X_t}
\circ dX_t + 3 T. 
\end{equation*}
We want to replace the innermost Stratonovich integral with its
backward counterpart, but the process $X_s$ is not backward adapted so we
cannot proceed straightforwardly. On the other hand by passing to the limit in
the Riemann sums it is easy to check that the following is appropriate:
\begin{equation*}
\tilde{W}_k =  
2 \real \int_0^T \left( \int_0^t e^{ik \cdot (X_t-X_s)} \circ \hat{d}X_s \right) 
\circ dX_t + 3 T
\end{equation*}
where now $X_t-X_s$ is $\mathcal{F}^t_s-$adapted and everything is well defined. 
Recall that the relation between backward Stratonovich and backward
It\^o integrals is
$$
\int_0^t f(Y_s)\circ \hat{d}X_s = \int_0^t f(Y_s)\hat{d}X_s -
\frac{1}{2} \int_0^t \nabla f(Y_s) \,ds
$$
for processes $(Y_s)_{s\in[0,t]}$ adapted to the backward filtration $\mathcal{F}^t_s$.

We have 
\begin{equation*}
\begin{split}
\tilde{W}_k & =  2 \real \int_0^T \left( \int_0^t e^{ik \cdot (X_t-X_s)} \hat{d}X_s \right)
\circ dX_t + 3T
\\ & \qquad + \real \int_0^T \nabla_{X_t}  \left( \int_0^t e^{ik \cdot
 (X_t-X_s)} ds \right)  \circ dX_t 
\end{split}
\end{equation*}
where $\nabla_{X_t}$ is the gradient on functions of $X_t$.
We can perform the last integral noting that if $\varphi(t,X_t)$
is a generic function of $t$ and $X_t$ then 
$$
d \varphi(t,X_t) = \nabla_{X_t} \varphi(t,X_t) \circ dX_t +
\frac{\partial}{\partial t}  \varphi(t,X_t)\, dt
$$
so
\begin{equation*}
\begin{split}
\tilde{W}_k & 
 =  
2 \real \int_0^T \left( \int_0^t e^{ik \cdot (X_t-X_s)} \hat{d}X_s \right) 
dX_t  
\\ & \qquad 
+ \real \int_0^T \nabla_{X_t} \left( \int_0^t e^{ik \cdot (X_t-X_s)} \hat{d}X_s \right) 
dt   \\ & \qquad + 2T + \real \int_0^T e^{ik \cdot
 (X_T-X_s)} ds 
\end{split}
\end{equation*}
where we have also transformed the first Stratonovich in It\^o plus correction
according to
$$
\int_0^T f(X_t)\circ dX_t = \int_0^T  f(X_t)\,dX_t + \frac{1}{2} \int_0^T  \nabla f(X_t)\, dt.
$$ 
Proceeding in the same way we can compute also the second stochastic
integral. First, we rewrite it in  Stratonovich form
\begin{equation*}
\begin{split}
& \real \int_0^T \nabla_{X_t} \left( \int_0^t e^{ik \cdot (X_t-X_s)}\, \hat{d}X_s \right) 
dt = \\ & \qquad 
- \real \int_0^T \left( \int_0^t \nabla_{X_s} e^{ik \cdot (X_t-X_s)} \circ \hat{d}X_s \right) 
dt 
\\ & \qquad  
- \frac{1}{2} \real \int_0^T \left( \int_0^t \Delta_{X_s} e^{ik \cdot (X_t-X_s)} ds \right) 
dt
\end{split}
\end{equation*}
which can now be trivially performed. At the end we have 
\begin{equation*}
\begin{split}
\tilde{W}_k &  =  
2 \real \int_0^T  \int_0^t e^{ik \cdot (X_t-X_s)} \, \hat{d}X_s  dX_t  
\\ & \qquad 
+ \real \int_0^T \left(e^{ik \cdot (X_T-X_t)} +  e^{ik \cdot
    (X_t-X_0)}\right)\, dt  
\\ & \qquad 
+ \frac{\|k\|^2}{2} \real \int_0^T \int_0^t e^{ik \cdot (X_t-X_s)}\, ds~dt + T \\
\end{split}
\end{equation*}
From this expression we can recover the formula~(\ref{equivalence}) recalling that
$$
H = \int d\nu(k) \tilde{W}_k.
$$
The proof is complete. 
\qed

\begin{corollary}
\label{corollary_of_equivalence}
If $(X_t)$ is the Brownian motion, then
\begin{equation}
\begin{split}
H & = \int_0^T \!\int_0^T  \tr \left[B^\rho(X_t-X_s)\, \hat{d}X_s \otimes
 dX_t \right]+ 
   G^\rho(0) T   \\
\end{split}
\label{equivalence2}
\end{equation}
where
$$
B^\rho(x) = \int_{\mathbb{R}^3} p_k \,e^{ik \cdot x} d\nu(k). 
$$
\end{corollary}

PROOF. 
It\^o formula gives
\begin{equation*}
W_k  \equiv  \left\| p_k Y_{k,T} \right\|^2 
 =  
2 \real \int_0^T \left( \int_0^t e^{-ik \cdot X_s} p_k dX_s \right) e^{ik \cdot X_t}
p_k dX_t + \tr (p_k) T
\end{equation*}
and following the trace of the proof of theorem~\ref{th:equivalence}
and using the fact that $\tr B^\rho(0) = G^\rho(0)$ the
result is easily proved.
\qed

\begin{remark}
Theorem~\ref{th:equivalence} cannot be straightforwardly extended to a Brownian
semimartingale since backward integrals are not properly defined. 

Nonetheless also in the general setting of the present work it is still
possible to show the presence of a regularized intersection local time.
With the notations of the last theorem, consider
\begin{equation*}
\begin{split}
\Tilde{W}_k & = \left\| \int_0^T e^{ik\cdot X_t} dX_t \right\|^2
+  \real \left(\int_0^T e^{-ik\cdot X_t}\, ik \cdot dX_t\right) \left(\int_0^T
  e^{ik\cdot X_t} dt\right) \\
& \qquad + \frac{\|k\|^2}{4} \real \int_0^T \int_0^T e^{ik\cdot (X_t-X_s)}\, dt~ds. 
\end{split}
\end{equation*}
From which we obtain easily the decomposition
\begin{equation*}
\begin{split}
\Tilde{W}_k & = \left\| \int_0^T e^{ik\cdot X_t} dX_t \right\|^2
+  \real \int_0^T \left( e^{ik\cdot (X_T-X_t)}-e^{ik\cdot (X_0-X_t)}\right)\, 
 dt \\
& \qquad + \frac{3 }{4} \|k\|^2 \real \int_0^T \int_0^T e^{ik\cdot (X_t-X_s)}\, dt~ds. 
\end{split}
\end{equation*}
In this expression the last term generates, after the $k$ integration, a term
proportional to the $(\rho*\rho)-$regularized intersection local time. 

Equation~(\ref{eq:finitedifference}) tells us that there is no real
difference between the two energies $H$ and $\Tilde{H}$, at least as
far as the singular behaviour connected to the presence of the
intersection local time is concerned. This is true in particular with
respect to the result of Corollary~\ref{corollary_of_equivalence}
where the intersection local time does not show up explicity.
\end{remark}

\section{Exponential bound}
As a direct consequence of the positivity and finiteness of the energy we
have:
\begin{corollary}
Let $\left( X_{t}\right) $ be a Brownian semimartingale on the
filtered probability space $( \Omega ,\mathcal{A},( \mathcal{F}%
_{t}) _{t\in \left[ 0,T\right] },\prob) $ of the form (\ref
{Browsemim}), with drift satisfying the condition of theorem~\ref{th1}. 
Let $\rho$ be a finite measure on $(\mathbb{R}^{3},B(\mathbb{R}^{3}))$ 
satisfying (\ref{rho}). Given a real number $\beta$ define the
partition functions 
\begin{equation*}
Z_{\beta }\equiv \expect e^{-\beta H}, \qquad
 \Tilde{Z}_{\beta }\equiv \expect e^{-\beta \Tilde{H}}
\end{equation*}
and the Gibbs measures 
\begin{equation*}
d\mu_{\beta }=\frac{1}{Z_{\beta }}e^{-\beta H}d\prob,
\qquad d\Tilde{\mu} _{\beta }=\frac{1}{\Tilde{Z}_{\beta }}e^{-\beta \Tilde{H}}d\prob.
\end{equation*}
Then for all $\beta \ge 0$ we have that
$$
\Tilde{Z}_\beta \le Z_\beta < \infty 
$$
and that the measures $\mu_\beta$ and $\Tilde{\mu}_\beta$
are well defined on $( \Omega ,\mathcal{A},\prob ) $.
\end{corollary}

Let us prove that, under additional assumptions on the drift $b$, the Gibbs
measures are well defined also for some negative temperatures. At the same
time we get the existence of the characteristic functions of
$\tilde{H}$ and $H$.

\begin{theorem}
\label{th_finite1}
Let $( X_{t}) $ be a Brownian semimartingale of the form (\ref
{Browsemim}), with drift satisfying, for some constants $M$ and $C_{b}^{\ast
}$, 
\begin{equation}
\expect \exp\left(\lambda \int_{0}^{T}\left\| b_{s}\right\| _{\mathbb{R}^{3}}ds\right)
\leq Me^{C_{b}^{\ast }\lambda ^{2}}\text{ for all }\lambda \geq 0.
\label{ass1}
\end{equation}
Let $\rho $ be a finite measure on $( \mathbb{R}^{3},B( 
\mathbb{R}^{3})) $ satisfying (\ref{rho}). Then there exists 
a real number $\gamma_* > 0$ (one can take $\gamma_*=1/(2AT)$
where $A$ is defined below in (\ref{defA})) such that 
\begin{equation*}
|\expect e^{z \Tilde{H}}| < \infty, \qquad   |\expect e^{z H} | <\infty, 
\end{equation*}
for all complex numbers $z$ such that $| z | < \gamma_*$. 

This implies that all the moments of $H$ and $\Tilde{H}$ are finite
and that the Gibbs measures 
$\mu _{\beta }$, $\Tilde{\mu}_\beta$ are well defined for all
temperatures $\beta \ge -\gamma_* $.
\end{theorem}

PROOF. We use the notation $d\nu(k)$ of theorem~\ref{th1}. For complex 
$z$ we have
\begin{equation*}
\expect \left| e^{z H}\right| \leq \expect e^{\left| z
H\right| }  = \expect e^{\left|z \right| H} 
\end{equation*}
so it is sufficient to prove the theorem for positive real $z$.
Moreover the proof of the statement for $\Tilde{H}$ follows from that
for $H$ noting that
from~(\ref{eq:extraction_in_difference}), we have that
\begin{equation}
\expect e^{\gamma (\Tilde{H}-H)}  \le \expect  e^{\gamma D \|X_T-X_0\| }  
\end{equation} 
which easily is proved finite for all $\gamma\ge 0$ using the assumption~(\ref{ass1}).

Let us now prove that $\expect e^{\gamma H}<\infty$ for sufficiently
small $\gamma \geq 0$. Define
$$
Y^W_{k,T} \equiv \int_0^T e^{i k\cdot X_t} dW_t, \qquad
 Y^b_{k,T} \equiv \int_0^T e^{i k\cdot X_t} b_t dt
$$
and
$$
H^W \equiv \int_{\mathbb{R}^3}d\nu(k) \|p_k Y^W_{k,T}\|^2, \qquad
H^b \equiv \int_{\mathbb{R}^3}d\nu(k) \|p_k Y^b_{k,T}\|^2,
$$
then, since
\begin{equation}
\begin{split}
H & = \int_{\mathbb{R}^3} d\nu(k) \|p_k Y^W_{k,T} + p_k Y^b_{k,T} \|^2 \\
& \le  2 \int_{\mathbb{R}^3}d\nu(k) \left( \|p_k Y^W_{k,T} \|^2 + \|p_k Y^b_{k,T}
  \|^2\right)^2 \\ & = 2 H^W + 2 H^b
\end{split}
\label{eq:pitagora}
\end{equation}
and
\begin{equation*}
\begin{split}
H^b & \le \int_{\mathbb{R}^3}d\nu(k)\left( \int_0^T \|b_t\| dt\right)^2 
= A \left( \int_0^T \|b_t\| dt\right)
\end{split}
\end{equation*}
where 
\begin{equation}
A \equiv \int d\nu \left( k\right) < \infty.
  \label{defA}
\end{equation}

We have, for every $\gamma \ge 0$, 
\begin{equation}
\begin{split}
\expect e^{ \gamma H } &\le \expect \left[ e^{  2 \gamma  H^W +  2 \gamma
  H^b  } \right] \le
 \left(\expect  e^{ 4 \gamma  H^b}\right)^{1/2}
\left(\expect  e^{ 4 \gamma  H^W}\right)^{1/2}
\\ & \le  \sqrt{M} e^{ 8 C^*_b A^2 \gamma^2}  \left(\expect  e^{ 4 \gamma  H^W}\right)^{1/2}. 
\end{split}
\label{splitting_hw_and_hb}
\end{equation}

By Jensen's
inequality we have
\begin{equation}
\begin{split}
\expect e^{4 \gamma H^W} \leq \int_{\mathbb{R}^3} 
\frac{d\nu \left( k\right)}{A} \expect 
\exp \left[ 4 \gamma A\|
  p_k Y^W_{k,T}\|^2 \right].
\end{split}
\label{mainjensen}
\end{equation}
The proof will follow if, for sufficiently small $\gamma$, we bound
the expectation~(\ref{mainjensen}) uniformly in $\|k\|$. 
We give two proof of this fact, i.e.
of the following statement: for sufficiently small $\gamma$ there exists a
constant $C_{\gamma }<\infty $ such that 
\begin{equation}
\expect\left[ e^{4 A\gamma \| p_k Y^W_{k,T}\|^{2}}\right] \leq C_{\gamma }\text{ for all }%
k\text{.}  \label{stima-base}
\end{equation}

For the first proof we show that
\begin{equation}
\prob\left( \| p_k Y^W_{k,T} \|^{2}\geq n\right) 
\leq 8 \exp \left[ -\frac{n}{8T}\right]
\label{LDP2}
\end{equation}
which can be understood as a sort of upper large deviation result: 
\begin{equation}
\lim \sup_{n\rightarrow \infty }\frac{1}{n}\log \prob\left( \|p_k Y^W_{k,T}\|^{2}\geq
n\right) \leq -\frac{1}{8T}  \label{LDP}.
\end{equation}
The estimate (\ref{stima-base}) easily follows from this one, since 
\begin{equation*}
\begin{split}
\expect\left[ e^{4 A\gamma \left\| p_k Y^W_{k,T}\right\|^{2}}\right]
& =\sum_{n=0}^{\infty }\expect\left[ e^{4 A\gamma \left\|
      p_k Y^W_{k,T}\right\|^{2}}
1_{\left\{ n\leq \left\| p_k Y^W_{k,T}\right\|^{2}<n+1\right\} }\right] 
\\ &\leq e^{4 A\gamma }\sum_{n=0}^{\infty }e^{4 A\gamma n}\prob\left( \left\|
p_k Y^W_{k,T}\right\|^{2}\geq n\right) 
\\ &\leq 8 e^{4 A\gamma }\sum_{n=0}^{\infty }e^{4 A (\gamma - 1/(8AT))
  n}. 
\end{split}
\end{equation*}

Let us prove~(\ref{LDP2}). Consider an orthonormal basis of
$(e_1,e_2,e_3)$ of $\mathbb{R}^3$ with $e_3 \| k$ and let
$Y_{k,T}^{W,i} \equiv \langle e_i, p_k Y_{k,T}^{W}\rangle$.
The event $\{\left\| p_k
  Y^W_{k,T}\right\|^{2}\geq n\}$ 
is included in the union of the events $\{(\real 
Y_{k,T}^{W,i})^{2}\geq n/4\}$ and $\{(\imag Y_{k,T}^{W,i})^{2}\geq n/4\}$, $i=1,2$%
. The event $\{(\real Y_{k,T}^{W,i})^{2}\geq n/4\}$ is included into the event $%
\{\real Y_{k,T}^{W,i}\geq \sqrt{n/4}\}\cup \{-\real Y_{k,T}^{W,i}\geq \sqrt{n/4}\}$,
and similarly for the others. We bound $\prob\left( \real Y_{k,T}^{W,i}\geq \sqrt{n/4}%
\right) $ by $\exp(-n/8T)$
and similarly for the other events. This implies (\ref{LDP2}). 

Let us check
the computation for one of these events. For every $\lambda \geq 0$ we have 
\begin{equation*}
\prob\left( -\real Y_{k,T}^{W,1}\geq \sqrt{\frac{n}{4}}\right)  =
\prob\left(
e^{-\lambda \real Y_{k,T}^{W,1}}\geq e^{\lambda \sqrt{\frac{n}{4}}}\right)  
\leq e^{-\lambda \sqrt{\frac{n}{4}}}\expect
e^{-\lambda \real Y_{k,T}^{W,1}}.
\end{equation*}
If we set $W^i_t \equiv \langle e_i,W_t\rangle$, we have
\begin{eqnarray*}
-\lambda \real Y_{k,T}^{W,1} 
&=&-\lambda \int_{0}^{T}\cos \left( k\cdot X_{t}\right) dW^1_{t}
\\
& \le &-\lambda \int_{0}^{T}\cos \left( k\cdot X_{t}\right) dW^1_{t}
-\frac{\lambda^{2}}{2} \int_{0}^{T}\cos ^{2}\left( k\cdot X_{t}\right) dt 
+\frac{\lambda ^{2}}{2} T.
\end{eqnarray*}
Therefore
\begin{eqnarray*}
&&\prob\left( -\real Y_{k,T}^{W,1}\geq \sqrt{\frac{n}{4}}\right)  \\
&\leq &e^{-\lambda \sqrt{\frac{n}{4}} + \lambda^2 T/2}
\expect\left[ e^{-\lambda \int_{0}^{T}\cos
\left( k\cdot X_{t}\right) dW^1_t 
-(\lambda ^{2}/2) \int_{0}^{T}\cos ^{2}\left( k\cdot X_{t}\right) dt}
\right]  \\
&\leq &e^{-\lambda \sqrt{\frac{n}{4}} +\lambda ^{2}T/2}.
\end{eqnarray*}
This last bound is minimized with respect to $\lambda$ by taking 
$\lambda =\sqrt{n}/2T$, which implies 
\begin{eqnarray*}
\prob
\left( -\real Y^{1,W}_{k,T} \geq \sqrt{\frac{n}{4}}\right)  &\leq &
\exp\left(-\frac{n}{8 T} \right).
\end{eqnarray*}
Note that the last computations are similar to the proof of Bernstein's
exponential inequality for martingales (see~\cite{R-Y}, Ch. IV, Ex.(3.16)).
This first proof is complete.

Another proof of~(\ref{stima-base})
exploits an auxiliary gaussian variable to simplify the estimation of
the stochastic integral.

Let us introduce a gaussian random variable $Z\in\mathbb{C}^{3}$,
independent from $(X_t)$, of mean zero and unit 
covariance ($\expect Z_{i}^{*} Z_{j} = \delta_{ij}$). 
Then it holds that 
\begin{equation}
\begin{split}
\expect \exp \left(  4 A\gamma \left\|
p_k Y^W_{k,T} \right\|^{2}
\right) & = 
\expect \exp\left[
  \sqrt{8 A \gamma} \real \left \langle Z,p_k Y^W_{k,T}
\right\rangle_{\mathbb{C}^3} \right] 
\\ &=
\expect  
\exp\left[  \sqrt{8 A \gamma} \int_{0}^{T} \real \left \langle p_k Z,
e^{ik\cdot X_{t}} dW_{t}
\right \rangle_{\mathbb{C}^3} \right] 
\end{split}
\label{gaussiantrick1}
\end{equation}
where $\langle \cdot,\cdot \rangle_{\mathbb{C}^3}$ is the scalar product in $\mathbb{C}^3$.

We can estimate
\begin{equation}
\begin{split}
& \expect\exp\left[  4 A\gamma \left\|
p_k Y^W_{k,T} \right\|^{2}
\right]  =
\expect  
\exp\left[  \sqrt{8 A \gamma} \int_{0}^{T} \left\langle
\real e^{-ik\cdot X_{t}} Z, p_k dW_{t}
\right\rangle_{\mathbb{R}^3} 
\right]
\\ & =  \expect   \exp\left[  
\sqrt{8 A \gamma} \int_{0}^{T} \left\langle
\real e^{-ik\cdot X_{t}} p_k Z, dW_{t}
\right\rangle_{\mathbb{R}^3} 
\right. \\ & \left. \qquad
-  4 A \gamma \int_0^T  
\left\| \real e^{-ik\cdot X_{t}} p_k Z \right\|^2 dt
+  4 A \gamma \int_0^T  
\left\| \real e^{-ik\cdot X_{t}} p_k Z \right\|^2 dt
\right]
\\ & \le  \expect   \exp\left[  
\sqrt{8 A \gamma} \int_{0}^{T} \left\langle
\real e^{-ik\cdot X_{t}} p_k Z, dW_{t}
\right\rangle_{\mathbb{R}^3} 
\right. \\ & \left. \qquad
-  4 A \gamma \int_0^T  
\left\| \real e^{-ik\cdot X_{t}} p_k Z \right\|^2 dt
+  4 A \gamma T \|p_k Z\|^2 
\right]
\\ & =  \expect   \exp\left[  
  4 A \gamma T \|p_k Z\|^2 
\right]
\end{split}
\label{gt_secondbound}
\end{equation}

If $\gamma < (8AT)^{-1}$ the expectation is finite and we 
obtain the bound
\begin{equation*}
\begin{split}
& \expect\exp\left[  4 A\gamma \left\|
Y_{k,T} \right\|^{2}
\right]   \le
\left( 1- 8 A \gamma T \right)^{-2}.
\end{split}
\end{equation*}
The second proof is also complete.

Finally, a remark on the constant $\gamma_*$ of the statement of the
theorem. In~(\ref{eq:pitagora}) we can use the inequality
$$
\| p_k Y^W_{k,T} + p_k Y^b_{k,T} \|^2 \le (1+\epsilon) \| p_k
Y^W_{k,T}  \|^2 + (1+\epsilon^{-1}) \| p_k Y^b_{k,T} \|^2,
$$
true for all $\epsilon > 0$. In~(\ref{splitting_hw_and_hb}) we can use
H\"older inequality with a suitable couple of conjugate exponents. In
this way it is sufficient to prove an inequality of the
form~(\ref{stima-base}) for $\expect \exp\left( (1+\epsilon)^2 A
  \gamma \|p_k Y^W_{k,T}\|^2 \right)$ with arbitrary small
$\epsilon$. The second proof of~(\ref{stima-base}) now requires that
$\gamma < (2(1+\epsilon)^2 AT)^{-1}$, so $\gamma^*$ can be chosen equal
to $(2AT)^{-1}$.  \qed

\begin{remark} 
If $Z$ is a Gaussian random variable taking values in a separable Banach space
$E$ with norm $\|\cdot \|_E$, then the inequality
\begin{equation*}
2 \lambda \|Z\|_E \leq
\varepsilon \lambda ^{2}+\frac{\|Z\|_E^2}{\varepsilon }
\end{equation*}
valid for all $\varepsilon$ (the l.h.s. is the minimum in 
$\varepsilon$ of the r.h.s.) implies that
\begin{equation}
\expect e^{\lambda \|Z\|_E} \le e^{\varepsilon \lambda^2/2}
\expect e^{\|Z\|_E^2/(2\varepsilon)}.
\label{dominance}
\end{equation}
By Fernique's theorem~\cite{Fer} for sufficiently large $\varepsilon$ the
expectation on the r.h.s. is finite and thus there are positive constants $M$ and
$C$ such that
\begin{equation*}
\expect e^{\lambda \|Z\|_E}  \le Me^{C\lambda ^{2}}
\end{equation*}
for all $\lambda \ge 0$.
In this context the assumption (\ref{ass1}) becomes natural if we think at $Z =
(b_s)_{s\in[0,T]}$ on the space $E=L^1([0,T],ds)$ and
$$
\|Z\|_E = \int_{0}^{T}\left\| b_{s}\right\| _{\mathbb{R}^{3}} ds
$$
compared to more naive assumptions like, for example,   
$$
\expect  \exp
\left( \lambda \|Z\|_E \right) %
 \leq Me^{C\lambda }.$$
 \end{remark}


\begin{example}
\label{example_of_brownian_bridge}
(Brownian Bridge) In particular, in the case of the Brownian Bridge, 
$b_t$ is a continuous Gaussian process on $[0,T)$ and, by the estimate
of example 1 on
$X_t$ as $t \to T$, we have that $(b_t)$ can be seen as a Gaussian random
variable taking values in the separable Banach space $L^1(0,T)$.
In this case theorem~\ref{th_finite1} applies.
\end{example}

\subsection{Upper bound on $\gamma$}

For completeness we will show that, at least in the case of the Brownian
motion, the partition functions $\Tilde{Z}_{\beta},
Z_\beta$ cannot be finite for arbitrary large negative $\beta$.

\begin{theorem}
Let $( X_{t}) $ be a Brownian motion 
and $\rho $ a finite measure on $( \mathbb{R}^{3},B( 
\mathbb{R}^{3})) $ satisfying (\ref{rho}). Then there exists 
a real number $\gamma^* > 0$ such that for all  $\beta < -\gamma^*$
\begin{equation*}
\Tilde{Z}_\beta =  Z_\beta = \infty. 
\end{equation*}
\end{theorem}

PROOF. 
The proof aim at an explicit expression for $\gamma^*$ which could be
compared with the constant $\gamma_*$ obtained in Theorem~\ref{th_finite1}.
We will use the same notations of Theorem~\ref{th_finite1}.


We want to show that $H$ can be bounded from below by the square of Gaussian
random variable. 

Consider the spherical decomposition of 
$\mathbb{R}^3 = \mathbb{R}_+ \times S_2$, denoting $(q,\hat{k}) \in
\mathbb{R}_+ \times S_2$ spherical coordinates such that $k =
q\hat{k}$,  let
$$
\overline{\rho}(q)^2 \equiv \int_{S_2}
\frac{d\hat{k}}{4 \pi} |\hat{\rho}(q\hat{k})|^2   ,
\qquad d\overline{\nu}(k) \equiv \half
\frac{|\overline{\rho}(\|k\|)|^2}{\|k\|^2} \frac{dk}{(2\pi)^3}
$$
and note that $\int d\nu(k) = \int d\overline{\nu}(k)$.

For every function $f \in L^2(\mathbb{R}^3,d\nu)$ set 
$$
\overline{f}(q) \equiv 
\int_{S_2} \frac{d\hat{k}}{4 \pi} f(q\hat{k})
  \frac{\hat{\rho}(q\hat{k})}{\overline{\rho}(q)}   
$$ 
which depend only $q = \|k\|$. 

We have
\begin{equation*}
\begin{split}
\int_{\mathbb{R}^3} d\nu(k) |f(k)|^2 & = 2 \pi \int_{\mathbb{R}_+}
\frac{dq}{(2\pi)^3}
\frac{\overline{\rho}(q)^2}{q^2} \int_{S_2} \frac{d\hat{k}}{4 \pi}
\left| f(q\hat{k}) \frac{\hat{\rho}(q\hat{k})}{\overline{\rho}(q)}\right|^2
\\ & \ge 2 \pi \int_{\mathbb{R}_+} \frac{dq}{(2\pi)^3}
\frac{|\overline{\rho}(q)|^2}{q^2} 
\left|\int_{S_2} \frac{d\hat{k}}{4 \pi} f(q\hat{k})
  \frac{\hat{\rho}(q\hat{k})}{\overline{\rho}(q)} \right|^2   
\\ & =  2 \pi \int_{\mathbb{R}_+} \frac{dq}{(2\pi)^3}
\frac{|\overline{\rho}(q)|^2}{q^2} 
|\overline{f}(q)| ^2  
= \int_{\mathbb{R}^3} d\overline{\nu}(k) |\overline{f}(\|k\|)|^2.
\end{split}
\end{equation*}

Then the following lower bound for $H$ holds:
\begin{equation}
\begin{split}
H  
& =
\int d\nu(k)
\left\| \frac{k}{\|k\|}  \wedge Y_{k,T}\right\|^{2} \\
& \ge
\int d\overline{\nu}(k)
\left\|\int_{0}^{T} \left( \int_{S_2} \hat{v} e^{i \|k\| \hat{v} \cdot X_t}
    \frac{d\hat{v}}{4\pi} \right) \wedge dX_t
\right\|^{2} 
\\
 & \ge
\int d\overline{\nu}(k)
\left\|\int_{0}^{T} \frac{\Psi(\|k\|\,\|X_t\|)}{\|X_t\|}  (X_t \wedge dX_t)
\right\|^{2} 
\label{eq:energyvsarea}
\end{split}
\end{equation}
where
$$
\Psi(z) \equiv \frac{\cos(z)}{z} - \frac{\sin(z)}{z^2}
$$
for $z \ge 0$ and $|\Psi(z)| < 1/2$ and the computation of the integral over
$S_2$ can be easily performed using the following observation. Let
\begin{equation*}
\begin{split}
U(X_t) \equiv \int_{S_2} \hat{v} e^{i \|k\| \hat{v} \cdot X_t}
    \frac{d\hat{v}}{4\pi} 
\end{split}
\end{equation*}
and note that the vector $U(X_t)$ must be parallel to $X_t$ 
since if $R$ is a rotation of $\mathbb{R}^3$ we have 
\begin{equation*}
\begin{split}
U(X_t) & = \int_{S_2} \hat{v} e^{i \|k\| \hat{v} \cdot X_t}
    \frac{d\hat{v}}{4\pi} =
 \int_{S_2} (R \hat{v}) e^{i \|k\| (R\hat{v}) \cdot X_t}
    \frac{d (R \hat{v})}{4\pi} \\
& = \int_{S_2} (R \hat{v}) e^{i \|k\| \hat{v} \cdot R X_t}
    \frac{d \hat{v}}{4\pi} 
= R U(R X_t) 
\end{split}
\end{equation*}
then since $R$ is arbitrary it must hold that $U(X_t) = u X_t$ where $u \in
\mathbb{C}$. Then we can compute
\begin{equation*}
\begin{split}
u \|X_T\|^2 & = X_t \cdot V(X_t) = \int_{S_2} \hat{v}\cdot X_t e^{i \|k\|
  \hat{v} \cdot X_t} \frac{d\hat{v}}{4\pi} 
\\ & = \frac{\|X_t\|}{4\pi} \int_0^{2\pi} d\varphi \int_{-1}^1 d\!\cos \theta\, 
 \cos \theta e^{i \|k\|\,
  \|X_t\| \cos \theta } 
\\ & = -i \|X_t\|\, \Psi(\|k\|\, \|X_t\|).
\end{split}
\end{equation*}
which readily gives~(\ref{eq:energyvsarea}). 

Now, we introduce the stochastic area for the process $X_t$ in the plane
$1,2$ as
$$
S_t \equiv \int_0^t X_s^1 dX_s^2 - X_s^2 dX_s^1
$$
and the magnitude of $X_t$ in the plane $1,2$
$$
r^2_t \equiv \|X_t^1\|^2 + \|X_t^2\|^2
$$
and we can estimate
\begin{equation*}
\begin{split}
H & \ge 
\int d\overline{\nu}(k)
\left|\int_{0}^{T} \frac{\Psi(\|k\|\,\|X_t\|)}{\|X_t\|}  
   dS_t
\right|^2 
\equiv
\int d\overline{\nu}(k)
\|J_{k,T}\|^2
\end{split}
\end{equation*}
Lemma~\ref{lemma:skew-product} below shows that there exist a standard
Brownian motion $B_t$ independent from $(r_t)$ and from $(\|X_t\|)$  such that 
\begin{equation}
J_{k,T} = \int_0^T  \frac{\Psi(\|k\|\,\|X_t\|)}{\|X_t\|} r_t dB_t.
\label{eq:representation}
\end{equation}

Using this representation it is possible to  condition on $(B_t)$ in Jensen's
inequality to obtain
\begin{equation*}
\begin{split}
\expect e^{\gamma H} & \ge  
\expect \exp \left( \gamma \int_{\mathbb{R}^3} d\overline{\nu}(k)  
\left| 
\int_0^T  \expect \left[ \frac{\Psi(\|k\|\,\|X_t\|)}{\|X_t\|} r_t  \Big | (B_t)
\right]  dB_t
\right|^2  \right) \\
 & \ge  
\expect \exp \left( \gamma \int_{\mathbb{R}^3} d\overline{\nu}(k)
\left| 
\int_0^T  \expect \left[  \frac{\Psi(\|k\|\,\|X_t\|)}{\|X_t\|} r_t  
\right]  dB_t
\right|^2  \right) \\
&  =   
\expect \exp \left( \gamma  
\int_0^T  \int_0^T  C(t,s) dB_t dB_s 
\right) \\
\end{split}
\end{equation*}
where
$$
C(t,s) \equiv   \int_{\mathbb{R}^3} d\overline{\nu}(k) 
\expect \left[  \frac{\Psi(\|k\|\,\|X_t\|)}{\|X_t\|} r_t
\right]
\expect \left[  \frac{\Psi(\|k\|\,\|X_s\|)}{\|X_s\|} r_s \right].
$$

A simpler form for $C$ is obtained exploiting the following observation. 
Let $\alpha_t$ be the angle of $X_t$ with the plane $1,2$, holds that
$$
|\cos \alpha_t| = \frac{r_t}{\|X_t\|}.
$$ 
and let $R$ be a rotation of $\mathbb{R}^3$. 
Being a 3-d Brownian motion, the process $(X_t)$ is invariant under $R$, that
is $( R X_t )$ has the same law as $( X_t )$, which means also that
$|\cos \alpha_t|$ has the same law that  $|\cos (\alpha_t + \theta)|$ where
$\theta$ is a fixed angle. Then we have
\begin{equation}
\begin{split}
\expect \left[  \frac{\Psi(\|k\|\,\|X_t\|)}{\|X_t\|} r_t
\right] 
& = \int_0^{2\pi} \frac{d\theta}{2\pi} \expect \left[ \Psi(\|k\|\,\|X_t\|) |\cos \alpha_t| \right] \\
& =  \int_0^{2\pi} \frac{d\theta}{2\pi} \expect \left[ \Psi(\|k\|\,\|X_t\|)
  |\cos (\alpha_t + \theta)| \right] \\
& = 
\int_0^{2\pi} |\cos \theta| \frac{d\theta}{2 \pi}  
\expect \Psi(\|k\|\,\|X_t\|)   \\
& =  \frac{2}{\pi} \expect \Psi(\|k\|\,\|X_t\|) \\
\end{split}
\end{equation}
and finally obtain
$$
C(t,s) =  \frac{4 }{\pi^2} \int_{\mathbb{R}^3} d\overline{\nu}(k)
 \expect \left[ \Psi(\|k\|\,\|X_t\|)
 \right] \expect \left[ \Psi(\|k\|\,\|X_s\|) \right].
$$

It is easy to see that $C$ is a trace class symmetric operator on
$L^2(0,T)$, then there exists a positive constant $\gamma^*$ and a
Gaussian random variable $Z \in \mathbb{R}$ of unit variance such that
$$
\expect e^{\gamma H} \ge  \expect e^{\gamma Z^2/2\gamma^*}
$$ 
which implies that 
$\expect e^{\gamma H} =  \infty$ 
when $\gamma > \gamma^*$, proving the theorem. Moreover using the fact that
$|\Psi(z)| < 1/2$ it is easy to see that $C$ is bounded and 
$$
\|C\|_{L^2} \le A /\pi^2
$$
which implies that $\gamma^* \ge \pi^2/(2AT)$. \qed

The proof is completed by the next lemma which justifies the
representation~(\ref{eq:representation}) recalling some facts connected with 
the skew-product representation of a 2-d Brownian motion (see \cite{R-Y} Chap. V).

Let 
\begin{equation*}
X_{t}^{\left( 1,2\right) } \equiv \left( X_{t}^{1},X_{t}^{2}\right)
,\,\,\,\,\,X_{t}^{\left( 1,2\right) \perp } \equiv \left(
-X_{t}^{2},X_{t}^{1}\right) .
\end{equation*}

\begin{lemma}
\label{lemma:skew-product}
The Brownian motion 
\begin{equation*}
B_{t} \equiv \int_{0}^{t}\frac{1}{r_{s}}X_{s}^{\left( 1,2\right) \perp }\cdot
dX_{s}^{\left( 1,2\right) }
\end{equation*}
is independent of the processes $\left( r_{t}\right) $ and $\left( \left\|
X_{t}\right\| \right) $. 

\end{lemma}

PROOF. By It\^{o} formula,
\begin{equation*}
dr_{t}^{2}=2r_{t}d\tilde{B}_{t}+dt
\end{equation*}
\begin{equation*}
d\left\| X_{t}\right\| ^{2}=2\left\| X_{t}\right\| d\hat{B}_{t}+dt
\end{equation*}
where $( \tilde{B}_{t} ) $ and $( \hat{B}_{t} ) $ are
the Brownian motions defined as
\begin{equation*}
\tilde{B}_{t}:=\int_{0}^{t}\frac{1}{r_{s}}X_{s}^{\left( 1,2\right) }\cdot
dX_{s}^{\left( 1,2\right) }
\end{equation*}
\begin{equation*}
\hat{B}_{t}:=\int_{0}^{t}\frac{1}{\left\| X_{s}\right\| }X_{s}\cdot dX_{s}.
\end{equation*}
The process $( B_{t},\tilde{B}_{t}) $ is a martingale, with 
$$
[ B,\tilde{B}] _{t}=0, \qquad [ B,B ] _{t}=t, 
\qquad [  \tilde{B},\tilde{B}] _{t}=t,
$$ 
so the processes $( B_{t} ) $
and $( \tilde{B}_{t} ) $ are independent Brownian motions. 
The process $( r_{t}^{2} ) $ is
the pathwise unique solution of the equation 
$$
dr_{t}^{2}=2\sqrt{r_{t}^{2}}d \tilde{B}_{t}+dt, \qquad r(0)=0
$$
 and as such it is
progressively measurable with respect to the filtration generated by $( 
\tilde{B}_{t} ) $. Therefore the processes $( r_{t}^{2} ) $,
and  $( r_{t} ) $ (since it is positive), and $(
B_{t} ) $ are independent. The proof for $( \| X_{t} \|) $ is similar. 
\qed

\section{Further comments}

\paragraph{Energy spectrum.}
The stochastic integrals involved in the spectral analysis of this paper
provide a formula for the energy spectrum of the fluid. It seems worth to
state it explicitly for future numerical or asymptotic investigations. Let
us recall the definition of the energy spectrum. If $\hat{u}\left(
  k\right)$ 
denotes the Fourier transform of the velocity field $u(x)$, 
given the statistical ensemble $\mu _{\beta }$ (with expectation $\mathbb{E}_{\beta }$), the
corresponding energy spectrum $E(q)$, $q\geq 0$  is defined
as 
\begin{equation*}
E(q) \equiv \mathbb{E}_{\beta }\left[ \int_{\left\| k \right\| = q}\left\| 
\hat{u}\left( k \right) \right\| ^{2}dk \right] .
\end{equation*}
then, a computation analogous to~(\ref{eq:inertial_energy}), gives the
formula
\begin{equation*}
E(q)=\frac{1}{q^{2}}\int_{\left\|k\right\| =q}\left| \hat{\rho}%
\left( k \right) \right| ^{2}\,\,\expect_{\beta }\left[ \left\| p_k
  Y_{k,T} \right\|^2 \right] dk.
\end{equation*}

\paragraph{More than one vortex.}
Consider a vorticity field made of a finite number $N$ of vortex
filaments $(X_{t}^{\left( 1\right) })$, ... , $(X_{t}^{\left( N\right) })$.
Assume we describe them by a Gibbs ensemble, where we take as a reference
measure on the $N$ filaments the product of $N$ copies of the Wiener measure
(so at the level of the reference measure the filaments are independent),
and we couple the filaments with the Gibbs weight 
\begin{equation*}
Z_{N,\beta }^{-1}\exp \left( -\beta H_{N}\right) \,\,\,\,\,\,\,\text{where }%
H_{N} \equiv \sum_{i,j=1}^{N}H_{ij}
\end{equation*}
and where $H_{ii}$ is the energy of the $i$-th filament, as described in the
previous part of this paper, while $H_{ij}$ for $i\neq j$ is the interaction
energy, described below in this remark. The aim of this and the following
paragraph is to say that: a) this Gibbs ensemble is well defined, when a
cross-section $\rho $ is incorporated into the model as in the previous
sections; b) the interaction energy $H_{ij}$ for $i\neq j$ is well defined
(and has finite moments of all orders) even without the cross section (i.e.
in the limit when $\rho $ tends to a Dirac measure); c) we do not know
whether the Gibbs ensemble is also well defined when $H_{ii}$ contains the
cross section while $H_{ij}$ for $i\neq j$ does not.

Let us now consider point a). 
Preliminary, let us remark that in the case of more than one filament
it is unrealistic to take, as $(X_{t}^{\left( i\right) })$, processes with
the same value at time $t=0$, like Brownian motions starting at the origin. 
The initial point of the filaments should be
possibly different among the various filaments, and possibly chosen at
random. It is very easy to define rigorously a randomized initial condition
for the processes $(X_{t}^{\left( i\right) })$ in order to take into account
the previous remark. We omit this description for shortness and treat the
condition at time $t=0$ as deterministic and given. 

Let $(X_{t}^{\left( 1\right) })$, ... , $(X_{t}^{\left( N\right) })$ be $N$
independent copies of 3-dimensional Brownian motion (possibly with positions
different from zero at time $t=0$). Assume for simplicity of exposition that
they are realized on $N$ copies of the Wiener space, and we consider them
jointly on the product space. We define the (interaction) energy as
\begin{equation*}
H_{nm}=\int_{\mathbb{R}^{3}}d\nu \left( k\right)  \left\langle
p_{k}Y_{k,T}^{\left( n\right) },p_{k}Y_{k,T}^{\left( m\right)
  }\right\rangle_{\mathbb{C}^{3}}
\,\,\,\,\,\,n,m=1,...,N
\end{equation*}
where 
\begin{equation*}
Y_{k,T}^{\left( n\right) }=\int_{0}^{T}e^{ik\cdot X_{t}^{\left( n\right)
}}dX_{t}^{\left( n\right) },\,\,\,\,n=1,...,N.
\end{equation*}
It is easy to see, as in the case of the self-energy, that this expression
is consistent with the physical description of the kinetic energy. We have
used the same cross section $\rho $ for all filaments just to shorten the
notations, but the general case is only notationally more cumbersome. We now
have
\begin{equation*}
H_{N}=\sum_{n,m=1}^{N}H_{nm}=\int_{\mathbb{R}^{3}}d\nu \left( k\right)
\sum_{n,m=1}^{N}\left\langle p_{k}Y_{k,T}^{\left( n\right)
},p_{k}Y_{k,T}^{\left( m\right) }\right\rangle _{\mathbb{C}%
^{3}}
\end{equation*}
\begin{equation*}
=\int_{\mathbb{R}^{3}}d\nu \left( k\right) \left\langle 
p_{k}\left( \sum_{n=1}^{N}Y_{k,T}^{\left( n\right) }\right) ,p_{k}\left(
\sum_{m=1}^{N}Y_{k,T}^{\left( m\right) }\right) \right\rangle
_{\mathbb{C}^{3}}
\end{equation*}
so it is clear that 
\begin{equation*}
H_{N}\geq 0.
\end{equation*}
Similarly we see that
\begin{equation*}
H_{nm}\leq \frac{1}{2}\int_{\mathbb{R}^{3}}d\nu \left( k\right) \left(
\left\| p_{k}Y_{k,T}^{\left( n\right) }\right\| ^{2}+\left\|
p_{k}Y_{k,T}^{\left( m\right) }\right\| ^{2}\right) =\frac{1}{2}\left(
H_{nn}+H_{mm}\right) 
\end{equation*}
so
\begin{equation*}
H_{N}\leq N\sum_{n=1}^{N}H_{nn}.
\end{equation*}
This proves that the Gibbs measures 
$$
d\mu _{N,\beta }=Z_{N,\beta }^{-1}\exp \left( -\beta H_{N}\right) dW_{N}
$$ (where $W_{N}$ denotes the product
measure of $N$ copies of the Wiener measure) are well defined for all
positive inverse temperatures $\beta $ and also for small negative ones.

Concerning the behaviour of this Gibbs ensemble as $N\rightarrow \infty $,
we think that it is possible to obtain a mean field result similar to the
one of~\cite{PLL - M}, when the interaction energy is rescaled with a
factor $1/N$ 
(this is postponed to a future paper). The main drawback of
this model with respect to the more idealized one of~\cite{PLL - M} is that
here we do not have local energy functionals and we cannot rewrite mean
values using Feynman-Kac type formulae. However, the propagation of
chaos seems to hold also here.

\paragraph{Independent vortices}
We discuss point b) of the previous paragraph in the case $N=2$ for notational
simplicity. Let $(X_{t})$ and $\left( Y_{t}\right) $ be two independent
3-D Brownian motions, with $t\in \left[ 0,T\right] $. Assume for
simplicity of exposition that they are defined on two copies of the Wiener
space, with expectations denoted respectively by $\mathbb{E}_{X}$ and $%
\mathbb{E}_{Y}$; then we consider the two processes on the product space.

We want to define, \textit{without the cross-section} $\rho $, the
interaction energy between the two vortex filaments $(x+X_{t})$ and $\left(
y+Y_{t}\right) $, where $x$ and $y$ are two given points of $\mathbb{R}^{3}$%
, possibly equal. The most difficult case is when $x=y$, so we restrict to
this case (recall that, in the opposite case when the two Brownian motions
coincide, the energy is finite for $x\neq y$, but it is infinite for $x=y$,
see~\cite{Fl2}). The definition we consider here, for an easier comparison
with~\cite{Fl2}, is
\begin{equation*}
H_{X,Y}=\frac{1}{4\pi} \int_{0}^{T}\left( \int_{0}^{T}\frac{1}{\left\| X_{t}-Y_{s}\right\| }%
\circ dY_{s}\right) \circ dX_{t}
\end{equation*}
where now there are no more difficulties of adaptedness: each one of the previous
iterated integrals contains only adapted processes with respect to the
integrator. Repeating the computations of theorem 4 above (in the case $\rho
=\delta $), it is not difficult to find the formula
\begin{equation}
\begin{split}
H_{X,Y} & = \frac{1}{4\pi}
\int_{0}^{T}\left( \int_{0}^{T}\frac{1}{\left\| X_{t}-Y_{s}\right\| }%
dY_{s}\right) dX_{t}
\\ & \qquad
- \frac{1}{4} \int_{0}^{T}\left( \int_{0}^{T}\delta \left(
X_{t}-Y_{s}\right) ds\right) dt
\\ & \qquad
+\frac{1}{8 \pi}\int_{0}^{T}\left( \frac{1}{\left\| X_{t}\right\| }-\frac{1}{%
\left\| X_{T}-Y_{t}\right\| }\right) dt. 
\end{split}
 \label{inter}
\end{equation}
Let us sketch the proof that all terms are square integrable random
variables. Let us first treat the second, more difficult term. The fact that
it is finite is well-known in the literature on the intersection local time
(see for instance that~\cite{We} and~\cite{LeG1} use it to
describe the self-intersection local time of a single Brownian motion, by a
cluster expansion). A brief explanation of its finiteness comes from
Tanaka-Rosen formula as in~\cite{Yo1}:
\begin{equation}
\begin{split}
2 \pi \int_{0}^{T} \int_{0}^{T} \delta\left(
X_{t}-Y_{s}\right) ds dt & =  - \int_{0}^{T} dX_t  \cdot \int_{0}^{T} ds \frac{X_t -
Y_s}{\| X_t - Y_s \|^3} 
\\ \qquad & - \int_{0}^{T} ds \left( \frac{1}{\| X_T - Y_s\|} - 
\frac{1}{\|X_0 - Y_s \|}
\right) 
\end{split}
\end{equation}
Due to this formula, the difficult term to
be estimated in the second moment of $\int_{0}^{T}  \int_{0}^{t}\delta
( X_{t}-Y_{s} ) ds  dt$ is
\begin{equation}
\mathbb{E}_{Y}\mathbb{E}_{X}\left| \int_{0}^{T}\left( \int_{0}^{t}\frac{%
X_{t}-Y_{s}}{\left\| X_{t}-Y_{s}\right\| ^{3}}ds\right) \cdot dX_{t}\right| ^{2}
\label{eq:moments_xx}
\end{equation}
which is bounded by
\begin{equation*}
\mathbb{E}_{Y}\mathbb{E}_{X}\int_{0}^{T}\left| \int_{0}^{t}\frac{1}{\left\|
X_{t}-Y_{s}\right\| ^{2}}ds\right| ^{2}dt.
\end{equation*}

Since it will be useful below we deduce the finiteness of this expectation
from the fact that there exists a
positive constant $\lambda$ such that
\begin{equation}
\expect_X \expect_Y \exp\left(\lambda \int_0^T \int_0^T \frac{dt\, ds}{\|X_t -
    Y_s\|^2} \right) < \infty.
\label{eq:finite_exp_lambda}
\end{equation}
Indeed, the following It\^o formula applies
$$
\log \|X_t - Y_T\| -  \log \|X_t\| = \int_0^T dY_s \frac{X_t -
  Y_s}{\|X_t - Y_s\|^2} + \frac{1}{2} \int_0^T \frac{ds}{\|X_t - Y_s\|^2} 
$$
calling 
$$
Z_t \equiv  \int_0^T \frac{ds}{\|X_t - Y_s\|^2}
$$
we have 
\begin{equation*}
\expect_X \expect_Y \exp\left(\lambda \int_0^T \int_0^T \frac{dt\, ds}{\|X_t -
    Y_s\|^2} \right) < 
\expect_X \int_0^T \frac{dt}{T}\expect_Y \exp\left(\lambda  T Z_t \right)
\end{equation*}
and using H\"older inequality  we can obtain the recursive bound
\begin{equation*}
\begin{split}
\expect_Y e^{\lambda T Z_t}
& = 
\expect_Y e^{ \lambda T (\log \|X_t - Y_T\| -  \log \|X_t\|) -\lambda T \int_0^T dY_s \frac{X_t -
  Y_s}{\|X_t - Y_s\|^2}}
\\ & \le \left[ \expect_Y \left(\frac{ \|X_t - Y_T\|}{ \|X_t\|}\right)^{2\lambda T}\right]^{1/2} 
\left[\expect_Y e^{  -2\lambda T \int_0^T dY_s \frac{X_t -
  Y_s}{\|X_t - Y_s\|^2}}\right]^{1/2}
\\ & \le \left[ \expect_Y \left(\frac{ \|X_t - Y_T\|}{ \|X_t
 \|}\right)^{2\lambda T}\right]^{1/2} 
\left[\expect_Y e^{  4(\lambda T)^2 Z_t }\right]^{1/4}
\end{split}
\end{equation*}
where the last inequality is obtained using again H\"older inequality after
having added and subtracted the appropriate compensator for the stochastic
integral in the exponent.
For $0 \le \lambda T\le 1/4$
$$
\expect_Y e^{\lambda T Z_t} \le 
\left[ \expect_Y \left(\frac{ \|X_t - Y_T\|}{ \|X_t
 \|}\right)^{2\lambda T}\right]^{1/2} 
\left[\expect_Y e^{ \lambda T Z_t }\right]^{1/4}
$$ 
which means that
\begin{equation*}
\begin{split}
\expect_Y e^{\lambda T Z_t} & \le 
\left[ \expect_Y \left(\frac{ \|X_t - Y_T\|}{ \|X_t
  \|}\right)^{1/2}\right]^{2/3} \\ & \le
 \left( \expect_Y  \frac{ \|X_t - Y_T\|}{ \|X_t
 \|}\right)^{1/3} \le C \| X_t \|^{-1/3}
\end{split}
\end{equation*}
Taking the expectation over $X$ and integrating over $t$ we obtain
$$
\expect_X \int_0^T \frac{dt}{T} \expect_Y e^{\lambda T Z_t} \le 
C \int_0^T \frac{dt}{T}   \expect_X  \| X_t \|^{-1/3} < \infty
$$
so that Eq.~(\ref{eq:finite_exp_lambda}) is justified.

Then easily follows that the expectation~(\ref{eq:moments_xx}) is bounded. 

For the first term of (\ref{inter}) we apply twice the isometry formula 
\begin{equation*}
\mathbb{E}_{Y}\mathbb{E}_{X}\left[ \left| \int_{0}^{T}\left( \int_{0}^{T}%
\frac{dY_{s}}{\left\| X_{t}-Y_{s}\right\| } \right) dX_{t}\right| ^{2}\right]
=\mathbb{E}_{Y}\mathbb{E}_{X}\int_{0}^{T}\int_{0}^{T}
 \frac{  ds \, dt}{\left\| X_{t}-Y_{s}\right\| ^{2}}. 
\end{equation*}
Then this expectation is also finite and a similar estimation shows that the
last term  of (\ref{inter}) has finite second moment.
Actually, using~(\ref{eq:finite_exp_lambda}) is straightforward to prove that
$H_{X,Y}$ has finite moments of all orders. 

However, as in~\cite{Fl2}, the estimates for $\mathbb{E}_{Y}\mathbb{E}%
_{X}\left| H_{X,Y}\right| ^{n}$ coming from such arguments are of the
form $n^{n}$, (due to the constants in Burkholder-Davis-Gundy theorem), and
we cannot infer the exponential integrability of $H_{X,Y}$. Since $%
H_{X,Y}$ is not necessarily positive (the interaction energy between
different parts of a vortex field may be positive or negative), we do
not have the finiteness  of $\mathbb{E}e^{-\beta H_{X,Y}}$ even for positive $%
\beta $.

At this point we would like to remark that the exponential integrability of
$H_{X,Y}$ is very much related to the
result~(\ref{eq:finite_exp_lambda}). Given that there exists a upper bound for
the constant $\lambda$ in (\ref{eq:finite_exp_lambda}), in the sense that for
large $\lambda$, the expectation~(\ref{eq:finite_exp_lambda}) is infinite,
then it is very unlikely that  the double stochastic integral in $H_{X,Y}$ could
be exponentially integrable at all.  
As a simple example of this phenomenon think to a Gaussian random variable $x$
with variance $y^2$ and let $y$ be a standard Gaussian variable, then $\expect
\exp (\lambda \text{Var}(x)) =  \expect \exp (\lambda y^2) < \infty$ only if $\lambda <
1/2$ and $\expect \exp (\gamma x^2) = \infty$ for all $\gamma >0$.

\paragraph{Smoother vortex filaments.} 
Consider a vorticity field concentrated on a smooth deterministic
curve $(\gamma _{t})_{t\in \left[ 0,T\right] }$, with a cross-section $\rho $
satisfying the assumptions of this paper. If we define the energy as 
\begin{equation*}
H=\int_{\mathbb{R}^{3}}d\nu \left( k\right) \left\|
p_{k}\int_{0}^{T}e^{ik\cdot \gamma _{t}}\dot{\gamma}_{t}dt\right\| ^{2},
\end{equation*}
we immediately see that
\begin{equation*}
H\leq \left( \int_{0}^{T}\left\| \dot{\gamma}_{t}\right\| dt\right)
^{2}\int_{\mathbb{R}^{3}}d\nu \left( k\right) <\infty .
\end{equation*}
So we obtain the following conclusion: under the assumption $\int_{\mathbb{R}%
^{3}}d\nu \left( k\right) <\infty $ for the cross section, both smooth and
Brownian-semimartingale filaments have finite energy. By an ideal
interpolation, one could conjecture that intermediate processes like
fractional Brownian motions with Hurst parameter $H\in \left( 1/2%
,1\right) $ have the same property. Until now, it has been proved~\cite{Ida}
only that this fact holds true under the more restrictive condition on the
cross section that $\int_{\mathbb{R}^{3}}\left| \hat{\rho}\left( k\right)
\right| ^{2}dk<\infty $, equivalent to the assumption that $\rho $ has an $%
L^{2}$ density with respect to Lebesgue measure. In fact, for smooth curves
we believe that even weaker assumptions on $\rho $ could be sufficient
(while for Brownian motion the assumption $\int_{\mathbb{R}^{3}}d\nu \left(
k\right) <\infty $ cannot be generalized, see~\cite{Fl2}). Therefore, the
case of processes like fractional Brownian motion requires a better
understanding.

\paragraph{Gaussian representation.}
The results of the present work lead to think that it is possible to
apply some standard techniques of statistical mechanics to the study of
vortex filaments. This paragraph is intended to be a very informal
discussion on such topics.

We can look at the energy $H$ as a positive quadratic form  on the
space of random vorticity fields $\xi(x)$
and at the Boltzmann weight
$\exp (-\beta H)$ for $\beta \ge 0$ as the characteristic function of
a Gaussian random field $\varphi$ (in some sense dual to $\xi(x)$)
defined on a different probability space (with expectation
$\mathcal{E}$) and with covariance $\beta H$, in the sense that
$$
\mathcal{E}e^{i \langle \varphi, \xi \rangle }  = e^{-\beta H }   
$$    
where $\langle \cdot, \cdot \rangle$ would be an appropriate duality. 

This kind representations are well known and very useful in the study of the
statistical mechanics of particles with two-body interactions since they
allow to use field-theoretical methods to prove convergence to the
thermodynamic limit or cluster properties of correlations functions
(see e.g.\cite{Simon}).  

Indeed the introduction of the random field $\varphi$ transforms the
problem of the self-interaction of the vortex filament in a problem
which resembles the motion of a particle in a random media since we
would have
$$
\exp\left({i \langle \varphi, \xi \rangle}\right) = 
\exp\left({i \int_0^T \varphi(X_t) \circ dX_t}\right)
$$ 
and we can think at this weight as something like the Girsanov exponent
for a random (complex) drift corresponding to $i\varphi$. Another
possibility is to consider the functional of $\varphi$
$$
e^{G(i \varphi)} = \expect e^{i \langle \varphi, \xi \rangle } 
$$
which contains all the relevant information since by
differentiation we can recover the marginals of the vorticity field:
$$
\mathcal{E} -i \frac{\delta }{\delta \varphi(x)} e^{G(i \varphi)} =
\mathcal{E} \expect \left[ \xi(x) e^{i \langle \varphi, \xi \rangle} \right]
= \expect_\beta  \xi(x) 
$$
and so on.
The functional $G(\varphi)$ is convex by the following easy
computation
\begin{equation*}
\begin{split}
e^{G(\varphi_1+\varphi_2)} & = \expect e^{ \langle \varphi_1, \xi
  \rangle +  \langle \varphi_2, \xi \rangle } \\ & \le 
\left( \expect e^{2  \langle \varphi_1, \xi \rangle}\right)^{1/2} 
\left( \expect e^{2  \langle \varphi_2, \xi \rangle}\right)^{1/2} 
\\ & = e^{G(2\varphi_1)/2+G(2\varphi_2)/2}
\end{split}
\end{equation*}
and would allow to study the mean-field limit of the model in the
sense that for a collection of $N$ identical and independent vortices
we have
$$
\expect e^{-\beta H_N} = \expect \mathcal{E} e^{i \langle \varphi, \sum_{i=1}^N \xi^{(i)}
  \rangle} = \mathcal{E}  \prod_{i=1}^N \expect e^{i \langle \varphi_1, 
  \xi^{(i)} \rangle} = \mathcal{E} e^{N G(i \varphi)}
$$  
where $\xi^{(i)}$ is the vorticity field of the $i$-th vortex. Of
course to have a sensible $N \to \infty$ limit we need to rescale the
energy with a factor $1/N$ which can be simply achieved by the substitution
$\varphi \to \varphi/\sqrt{N}$. Then we are led to study the limiting
(stochastic) functional
$$
e^{G_{\text{mf}}(i \varphi)} = \lim_{N \to \infty} e^{N G(i \varphi / \sqrt{N})}.
$$
under the law of $\varphi$.
In~\cite{PLL - M} there is the proof that an analogous limit for
nearly-parallel vortices exists and delivers what can
be considered somehow as a field dual to the mean-field density of vorticity.     

\paragraph{Renormalization of the energy.} Another interesting problem
is the study of the possible limits of the Gibbs ensemble $\mu_\beta$
when the distribution $\rho$ tends to a Dirac mass. This limit
certainly requires a renormalization of the energy. The situation is
actually very similar to that of the construction of the three dimensional
polymer measure~\cite{Bo}~\cite{We} where indeed the limit exists and is
singular with respect to the reference Wiener measure. In the case of
the Brownian motion, the
decomposition proved in~\cite{Fl2} and recalled in
Section~\ref{sec:otherexp} suggest that the relevant divergencies
are similar to that of the intersection local time and only the double
stochastic integral seems to contains terms which do not have counterpart in the
polymer case. Nonetheless the arguments of~\cite{Bo}~\cite{We} cannot
be straightforwardly extended since in the proof of the
renormalizability of the polymer measure the positivity of the
interaction energy is fundamental while the interaction energy between
different parts of a vortex filament is not positive.

\subsection*{Acknowledgment} 
This work has been partially supported by the MURST project "Stochastic
Processes with Spatial Structure". 

\end{document}